\documentclass[sn-mathphys]{sn-jnl}

\usepackage{graphicx}
\usepackage{comment}
\usepackage{epstopdf}
\usepackage{hyperref}

\newcommand{\be}{\begin{equation}}
\newcommand{\ee}{\end{equation}}
\newcommand{\beq}{\begin{equation}}
\newcommand{\eeq}{\end{equation}}
\newcommand{\bea}{\begin{eqnarray}}
\newcommand{\eea}{\end{eqnarray}}
\newcommand{\ba}{\begin{array}}
\newcommand{\ea}{\end{array}}
\newcommand{\nn}{\nonumber}
\newcommand{\bi}{\begin{itemize}}
\newcommand{\ei}{\end{itemize}}
\newcommand{\bn}{\begin{enumerate}}
\newcommand{\en}{\end{enumerate}}
\newcommand{\bc}{\begin{center}}
\newcommand{\ec}{\end{center}}

\newcommand{\MP}{{M_{\rm P}}}

\begin{document}

\title[ On the number of $e$-folds in the Jordan and Einstein frames]{\centering  On the number of $e$-folds\\ in the Jordan and Einstein frames}

\author*[1]{\fnm{Antonio} \sur{Racioppi}}\email{antonio.racioppi@kbfi.ee}

\author[2]{\fnm{Martin} \sur{Vasar}}

\affil*[1]{\orgname{National Institute of Chemical Physics and Biophysics}, \orgaddress{\street{R\"avala 10}, \city{Tallinn}, \postcode{10143}, \country{Estonia}}}

\affil[2]{\orgdiv{Institute of Physics}, \orgname{University of Tartu}, \orgaddress{\street{W. Ostwaldi 1}, \city{Tartu}, \postcode{50411}, \country{Estonia}}}

\abstract{We investigate how inflationary predictions are affected by the difference in the number of $e$-folds between the Jordan and Einstein frames. We study several test models in relation to a Jordan frame defined by the common Higgs-inflation-like non-minimal coupling to gravity and consider two different formulations of gravity: metric and Palatini. We find that the difference is quite contained in case of a metric Jordan frame while can be quite remarkable in case of a Palatini Jordan frame. We also discuss a way to overcome the discrepancy by using a frame invariant physical distance and a consequently frame invariant number of $e$-folds.}


\maketitle

\section{Introduction} \label{sec:Introduction}

Several observations of the cosmic microwave background radiation (CMB) indicate that at large scales the Universe is flat and homogeneous. These properties can be explained by assuming an accelerated expansion during the very early Universe \cite{Starobinsky:1980te,Guth:1980zm,Linde:1981mu,Albrecht:1982wi}. Such an inflationary era is also able to generate and preserve the primordial inhomogeneities which generated the subsequent large-scale structure that we observe. In its simplest version, inflation is usually formulated by adding to the Einstein-Hilbert action one scalar field, the inflaton, whose energy density drives the near-exponential expansion.

Recently, the BICEP/Keck Array experiment \cite{BICEP:2021xfz} has reduced even more the available parameter space, disfavoring several inflationary models. On the other hand, two of the most popular models, namely Starobinsky~\cite{Starobinsky:1980te} and non-minimal Higgs inflation~(e.g. \cite{Bezrukov:2013fka}) still lie in the allowed region. Both of these two models share the fact that they are initially formulated in the Jordan frame, where gravity is non-mimimal, and then usually moved to the Einstein frame where the inflaton is minimally coupled to gravity and inflationary computations are simpler. 
However, the equivalence of the Einstein and Jordan frames (specially at the quantum level) is still a debated question (e.g.~\cite{Jarv:2014hma,Kuusk:2015dda,Kuusk:2016rso,Flanagan:2004bz,Catena:2006bd,Barvinsky:2008ia,DeSimone:2008ei,Barvinsky:2009fy,Barvinsky:2009ii,Steinwachs:2011zs,Chiba:2013mha,George:2013iia,Postma:2014vaa,Kamenshchik:2014waa,Miao:2015oba,Buchbinder:1992rb,Burns:2016ric,Fumagalli:2016sof,Bezrukov:2017dyv,Karam:2017zno,Karamitsos:2017elm,Narain:2017mtu,Ruf:2017xon,Ohta:2017trn,Ferreira:2018itt,Rinaldi:2018qpu,Karam:2018squ,Azri:2018gsz,Finn:2019aip,Francfort:2019ynz,Akin:2020mcr}.) 
On the other hand it has been proven (e.g \cite{Chiba:2008ia,Chiba:2008rp,Jarv:2016sow} and refs. therein) that it is possible to consistently perform frame invariant inflationary slow-roll computations in either the Jordan or the Einstein frame. Nevertheless, the definition of the number of $e$-folds (e.g. \cite{Karam:2017zno,Lerner:2009na} and refs therein), which is a quantity that measures the expansion of the Universe and, in its current definition, a clearly frame-dependent quantity, remains an unsolved issue.
Using the formalism of \cite{Jarv:2016sow}, it can be proven \cite{Karam:2017zno} that both the Jordan frame number of $e$-folds and the Einstein frame one can be written in terms of invariant quantities and that the difference is an invariant itself. Therefore the issue becomes a choice between two different invariants. However the ambiguity still remains.

Another  issue  that  arises  in the context of  non-minimally  coupled  theories concerns the \emph{choice} of the dynamical degrees of freedom.  In  the  metric formulation, the  metric is the only dynamical degree of freedom and the connection chosen to be the Levi-Civita one.  On the contrary,  in the Palatini formulation, the metric and the connection are treated as independent variables with their corresponding equations of motion. When the action is linear in the curvature scalar, the two formalisms  lead to equivalent theories, otherwise the theories are completely different \cite{Bauer:2008zj} and lead to different phenomenological results, as recently investigated in (e.g. \cite{Koivisto:2005yc,Tamanini:2010uq,Bauer:2010jg,Rasanen:2017ivk,Tenkanen:2017jih,Racioppi:2017spw,Markkanen:2017tun,Jarv:2017azx,Racioppi:2018zoy,Kannike:2018zwn,Enckell:2018kkc,Enckell:2018hmo,Rasanen:2018ihz,Bostan:2019uvv,Bostan:2019wsd,Carrilho:2018ffi,Almeida:2018oid,Takahashi:2018brt,Tenkanen:2019jiq,Tenkanen:2019xzn,Tenkanen:2019wsd,Kozak:2018vlp,Antoniadis:2018yfq,Antoniadis:2018ywb,Gialamas:2019nly,Racioppi:2019jsp,Rubio:2019ypq,Lloyd-Stubbs:2020pvx,Das:2020kff,McDonald:2020lpz,Shaposhnikov:2020fdv,Enckell:2020lvn,Jarv:2020qqm,Gialamas:2020snr,Karam:2020rpa,Gialamas:2020vto,Karam:2021wzz,Karam:2021sno,Gialamas:2021enw,Annala:2021zdt,Racioppi:2021ynx,Cheong:2021kyc,Mikura:2021clt,Ito:2021ssc,Antoniadis:2021axu}). 

This article is organized as follows.
In section~\ref{sec:theo}  we set the theoretical framework and introduce the difference between the metric and Palatini formulation and the difference between the Jordan and Einstein frame number of $e$-folds.
In section~\ref{sec:numerical}  we test numerically the issue of the number of $e$-folds for several inflationary models in relation to a Jordan frame defined by the common Higgs-inflation-like non-minimal coupling to gravity, considering both the metric and the Palatini formulation\footnote{A similar study was already performed in \cite{Karam:2017zno}. However such a study presents a numerical analysis only for a specific model and only for the metric formulation. According to our knowledge, it is the first time that such an analysis is performed for the Palatini formulation of non-minimal gravity.}.
In section \ref{sec:solution}  we discuss a way to overcome the discrepancy by introducing a frame invariant physical distance and a consequently frame invariant number of $e$-folds. 
Finally, in section \ref{sec:conclusions} we present our conclusions.

\section{Theoretical framework} \label{sec:theo}
We start by assuming the following Jordan frame action
\begin{equation}
S = \!\! \int \!\! d^4x \sqrt{-g^J}\left(-\frac{M_P^2}{2}f(\phi)R^J(\Gamma) + \frac{(\partial \phi)^2}{2}  - V(\phi) \right) ,
\label{eq:JframeL}
\end{equation}
where $M_P$ is the reduced Planck mass, $R^J$ is the Jordan frame Ricci scalar constructed from a connection $\Gamma$, $f(\phi)$ is the non-minimal coupling to gravity and $V(\phi)$ is the potential of the inflaton scalar. 
It is possible to perform the inflationary analysis in the Jordan frame, however, since cosmological perturbations are invariant under frame transformations (e.g \cite{Prokopec:2013zya,Jarv:2016sow}), it is usually convenient to move the problem to the Einstein frame and perform the analysis there. Such a frame is obtained via the Weyl transformation
\begin{eqnarray}
\label{eq:gE}
g^E_{\mu \nu} = f(\phi) \ g^J_{\mu \nu} \, , 
\end{eqnarray}
that leads to the Einstein frame action
\begin{equation}
S = \!\! \int \!\! d^4x \sqrt{-g^E}\left(-\frac{M_P^2}{2} R^E + \frac{(\partial \chi)^2}{2}  - U(\chi) \right) ,
\label{eq:L:classic}
\end{equation}
where the scalar potential $U(\chi)$ is given by
\beq
U(\chi) = \frac{V(\phi(\chi))}{f^{2}(\phi(\chi))} \, .
\label{eq:U:general}
\eeq
The canonically normalized field $\chi$ depends on the function $f(\phi)$ and on the assumed gravity formulation.
In the common metric formulation we have
\begin{equation}
\frac{\partial \chi}{\partial \phi} = \sqrt{\frac{3}{2}\left(\frac{M_P}{f}\frac{\partial f}{\partial \phi}\right)^2+\frac{1}{f}} 
  \label{eq:dphim}
\end{equation}
where the first term comes from the transformation of the Jordan frame Ricci scalar and the second from the rescaling of the Jordan frame scalar field kinetic term. 
On the other hand, in the Palatini formulation \cite{Bauer:2008zj}, the field redefinition is generated only by the rescaling of the inflaton kinetic term i.e.
\begin{equation}
\frac{\partial \chi}{\partial \phi} = \sqrt{\frac{1}{f}} 
  \label{eq:dphiP}
\end{equation}
where no additional term comes from the Jordan frame Ricci scalar\footnote{More details about change of frame and the Palatini formulation are given in Appendix \ref{appendix}.}.
Assuming slow-roll, the inflationary dynamics can described in the Einstein frame by the potential slow-roll parameters 
\beq
\epsilon = \frac{1}{2}M_{\rm P}^2 \left(\frac{1}{U}\frac{{\rm d}U}{{\rm d}\chi}\right)^2 \,, \quad
\eta = M_{\rm P}^2 \frac{1}{U}\frac{{\rm d}^2U}{{\rm d}\chi^2} \,.
\ee
Inflation takes place when $\epsilon \ll 1$. The consequent expansion of the Universe is measured in number of $e$-folds
\beq
N_e^E = \frac{1}{M_{\rm P}^2} \int_{\chi_f}^{\chi_N} {\rm d}\chi \, U \left(\frac{{\rm d}U}{{\rm d} \chi}\right)^{-1},
\label{eq:Ne}
\ee
where the field value at the end of inflation, $\chi_f$, is determined by $\epsilon(\chi_f) = 1$.  
The field value $\chi_N$ at the time a given scale left the horizon is given by the corresponding $N_e$. 
Other two important observables, i.e. the scalar spectral index and the tensor-to-scalar ratio are respectively written in terms of the slow-roll parameters as
\bea
n_s &\simeq& 1+2\eta-6\epsilon \label{eq:ns} \\
r &\simeq& 16\epsilon \, \label{eq:r} .
\eea
To reproduce the correct amplitude for the curvature power spectrum, the potential has to satisfy \cite{Planck2018:inflation}
\beq
\label{eq:As:constraint}
\ln \left(10^{10} A_s \right) = 3.044 \pm 0.014   \, ,
\ee
where
\beq
 A_s = \frac{1}{24 \pi^2 \MP^4} \frac{U(\chi_N)}{\epsilon(\chi_N)} \label{eq:As} \, .
\ee
This constraint is commonly used to fix the normalization of the inflaton potential.
When an exact solution for the inverse field redefinition $\phi(\chi)$ (and the Einstein frame potential $U(\chi)$) is not possible, all the phenomenological parameters can be still derived using $\phi$ as computational variable, the chain rule $\frac{\partial}{\partial \chi} = \frac{\partial}{\partial \phi} \frac{\partial \phi}{\partial \chi}$ and eq. \eqref{eq:dphim} or \eqref{eq:dphiP}. 

Even though we presented only the Einstein frame's equations, $r$, $n_s$ and $A_s$ can be unequivocally defined also in the Jordan frame and the original frame of definition should not affect the phenomenological results (e.g \cite{Prokopec:2013zya,Jarv:2016sow}). 
On the other hand, there is a subtle issue regarding the evaluation of  the number of $e$-folds and consequently $\chi_N$ and $\phi_N$. Assuming a FRW metric in the Jordan frame, the corresponding line element is
\begin{eqnarray}
 {\rm d} s_J^2 &=& {\rm d} t^2 - a_J^2(t) {\rm d} \bf x^2
\end{eqnarray}
where $a_J$ is the scale factor of the Jordan frame metric $g^J_{\mu\nu}$. Therefore we can define the number of $e$-folds in the Jordan frame as
\be
 N_e^J = \ln\left[ \frac{a_J(t_f)}{a_J(t_N)} \right] \, ,
\ee
where $t_f$ is the time when inflation ends (i.e. $\phi=\phi_f$ and $\chi=\chi_f$) and $t_N$ is the time when $\phi=\phi_N$ and $\chi=\chi_N$.
 Analogously, the Einstein frame line element coming from the scaling in eq. \eqref{eq:gE} is 
\begin{eqnarray}
 {\rm d} s_E^2 &=& {\rm d} t_E^2 - a_E^2(t) {\rm d} \bf x^2 \,
\end{eqnarray}
where we have defined
\be
 {\rm d} t_E = \sqrt{f} \, {\rm d} t \quad, \qquad  a_E(t) = \sqrt{f} \, a_J(t) \, , \label{eq:tE:aE}
\ee
and the corresponding number of $e$-folds in the Einstein frame is
\be
 N_e^E = \ln\left[ \frac{a_E(t_f)}{a_E(t_N)} \right] .
\ee
By using eq. \eqref{eq:tE:aE} and a bit of algebra we obtain
\bea
 N_e^E &=& \ln  \left[ \frac{a_E(t_f)}{a_E(t_N)} \right] = \ln  \left[ \frac{\sqrt{f(\phi_f)} a_J(t_f)}{\sqrt{f(\phi_i)} a_J(t_N)} \right] = \ln\left[ \frac{a_J(t_f)}{a_J(t_N)} \right] + \ln\left[ \sqrt{\frac{f(\phi_f)}{f(\phi_i)}} \right] \nn\\
 &=& N_e^J + \frac{1}{2} \ln\left[ \frac{f(\phi_f)}{f(\phi_i)} \right]
\eea
i.e. a mismatch between the two number of $e$-folds. This has a tremendous impact in the determination of $\chi_N$ and $\phi_N$ and therefore in the computation of the observables $r$, $n_s$ and $A_s$. Using the formalism of \cite{Jarv:2016sow}, it can be proven \cite{Karam:2017zno} that both $N_e^J$ and $N_e^E$ can be written in terms of invariant quantities and that the difference $\Delta N_e = N_e^E - N_e^J$ is an invariant itself. Therefore the issue is actual about a choice between two invariants. For practical purposes, a solution is found by invoking the slow-roll approximation, treating $\Delta N_e$ as subdominant under slow-roll and therefore ignoring it. Therefore full unequivocal invariance is restored as an approximated result under slow-roll with $N_e^E \simeq N_e^J$.  However, first of all, given the increase in precision of observational data, such a difference might play a key role in the actual compatibility of inflationary predictions with current and future data. Moreover, from a theoretical point of view, this does not solve the issue of finding an unequivocal definition for the number of $e$-folds and undermines all the idea behind frame equivalence.

\section{Numerical examples} \label{sec:numerical}

In this section we study the impact of the different definitions of the number of $e$-folds with some numerical examples. It is customary to set the theory in the Jordan frame and then move it to the corresponding Einstein one. In our case we act in the opposite way. We first set the theory in the Einstein frame and then we move it to corresponding Jordan frame, according to the gravity formulation under consideration, metric or Palatini. Therefore, we first consider the following Einstein frame inflaton potentials:

\begin{eqnarray}
 U(\chi)&=&\Lambda^4 \, \left(\frac{\chi}{\MP}\right)^n \hspace{2.7cm} \text{(power-law potential)}, \label{eq:powerlaw}\\
 U(\chi)&=&\Lambda^4 \left [ 1 + \cos\left(\frac{\chi}{\mu}\right) \right] \hspace{1.7cm} \text{(natural inflation)}, \label{eq:natural}\\
 U(\chi)&=&\Lambda^4 \left[ 1 - \left(\frac{\chi}{\mu}\right)^4 \right] \hspace{2cm} \text{(quartic hilltop inflation)}, \label{eq:hilltop} \\
 %
 U(\chi)&=&\Lambda^4 \left[ \tanh^2\left(\frac{\chi}{\sqrt{6\alpha}M_p}\right) \right] \qquad \text{($\alpha$ attractor)}, \label{eq:attractor}
\end{eqnarray}
%
and compute for each one of them inflationary predictions for $N_e^E\in[50,60]$. Then we compute the same prediction for $N_e^J\in[50,60]$ with the Jordan frame corresponding to the popular non-minimal coupling
\be
 f(\phi) = 1 + \xi \left(\frac{\phi}{\MP}\right)^2 \, . \label{eq:f}
\ee
\begin{figure}[t]
\begin{center}
    \includegraphics[width=\textwidth]{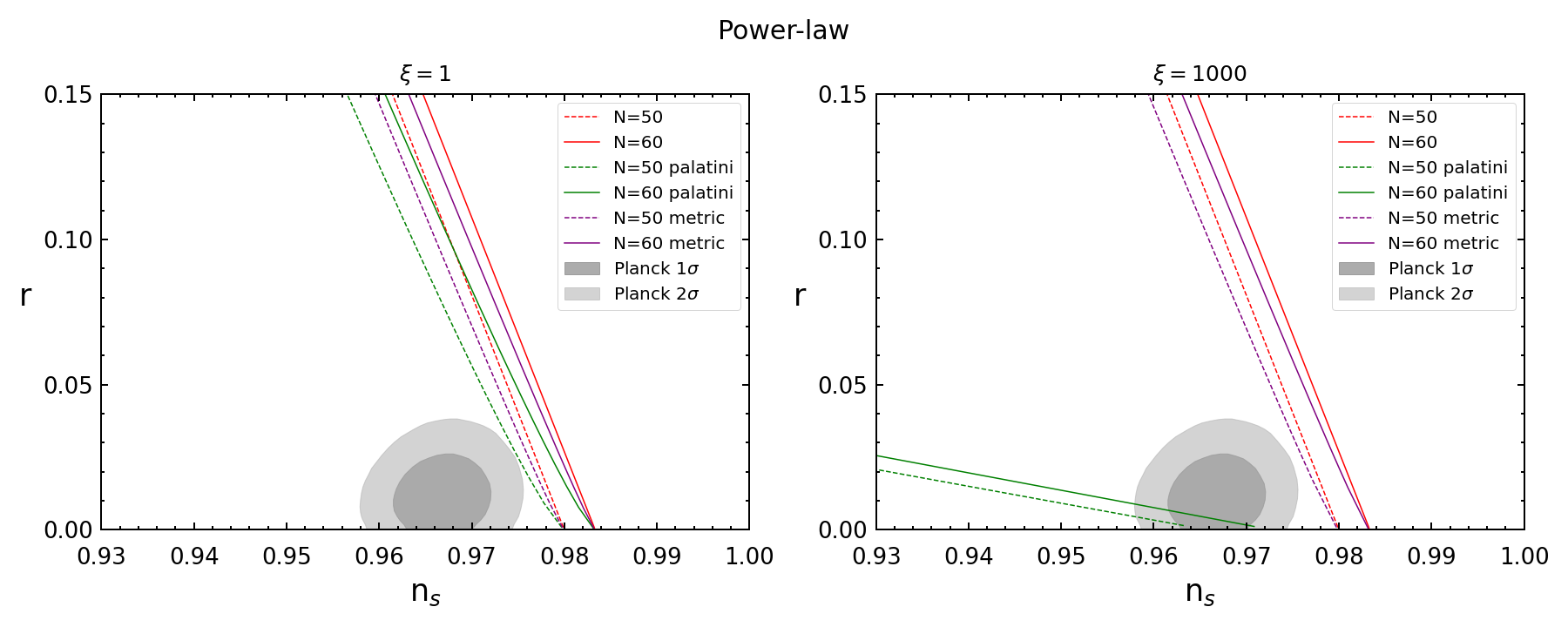} 
\end{center}
\caption{$r$ vs. $n_s$ for the power-law potential in eq. \eqref{eq:powerlaw} with $-2 < \log_{10} (n) < 2$ computed in the Einstein frame (red), in the metric Jordan frame (purple) and in the Palatini Jordan frame (green). Dashed (continuous) line represents $N_e=50\ (60)$. The gray areas represent the $1,2\sigma$ allowed regions from the latest combination of Planck, BICEP/Keck and BAO data \cite{BICEP:2021xfz}. }
\label{fig:powerlaw}
%
\begin{center}
    \includegraphics[width=\textwidth]{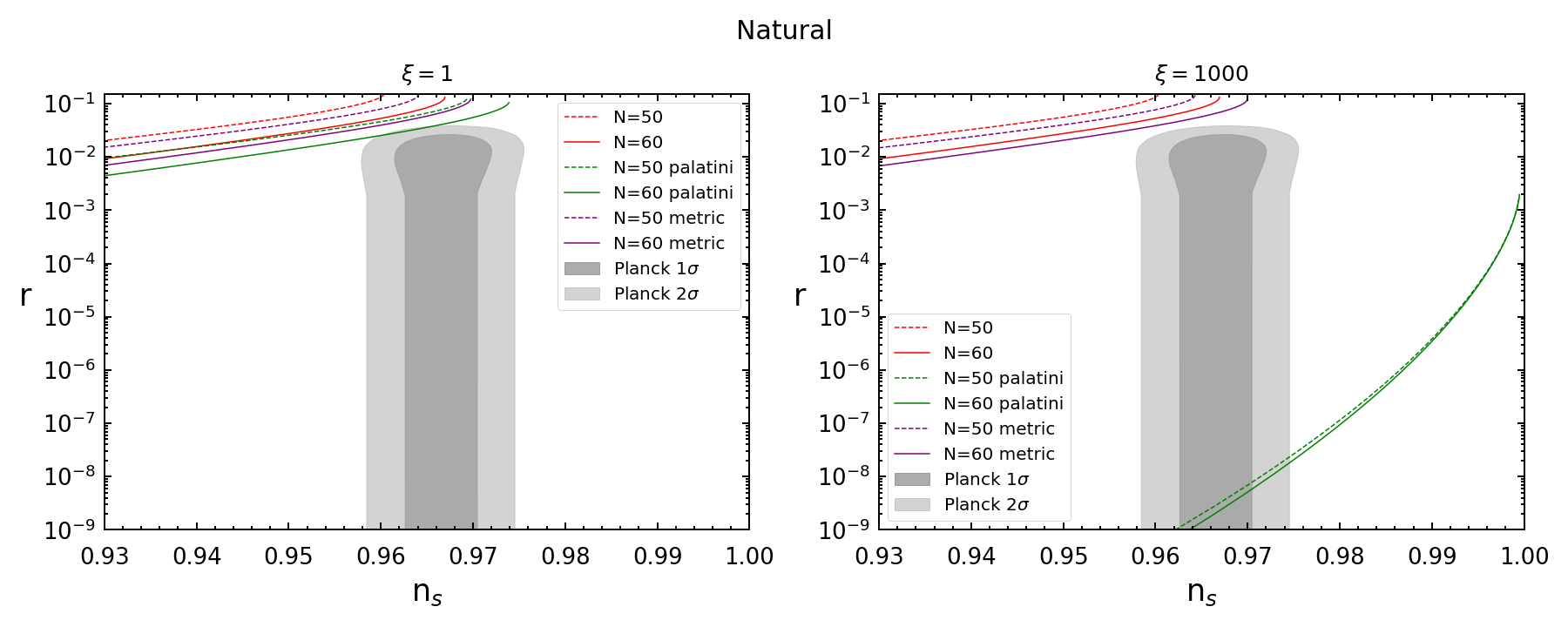}
\end{center}
\caption{$r$ vs. $n_s$ for the natural inflation potential in eq. \eqref{eq:natural} with $0 < \log_{10} (\mu/M_p) < 2.5$ computed in the Einstein frame (red), in the metric Jordan frame (purple) and in the Palatini Jordan frame (green). Dashed (continuous) line represents $N_e=50\ (60)$. The gray areas represent the $1,2\sigma$ allowed regions from the latest combination of Planck, BICEP/Keck and BAO data \cite{BICEP:2021xfz}. 
}
\label{fig:natural}
\end{figure}

\begin{figure}[t]
\begin{center}
    \includegraphics[width=\textwidth]{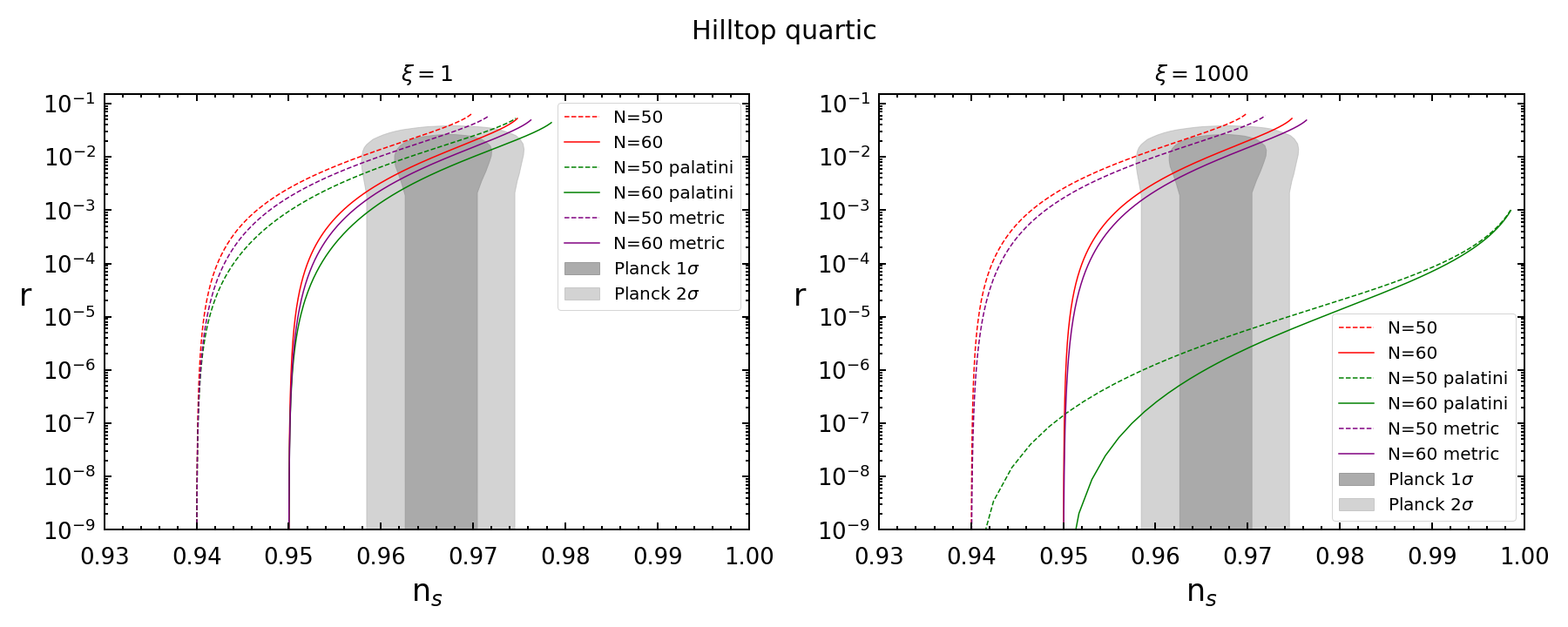}
\end{center}
\caption{$r$ vs. $n_s$ for the quartic hilltop potential in eq. \eqref{eq:hilltop} with $-2 < \log_{10} (\mu/M_p) < 2$ computed in the Einstein frame (red), in the metric Jordan frame (purple) and in the Palatini Jordan frame (green). Dashed (continuous) line represents $N_e=50\ (60)$. The gray areas represent the $1,2\sigma$ allowed regions from the latest combination of Planck, BICEP/Keck and BAO data \cite{BICEP:2021xfz}. 
}
\label{fig:Hilltop}
\begin{center}
    \includegraphics[width=\textwidth]{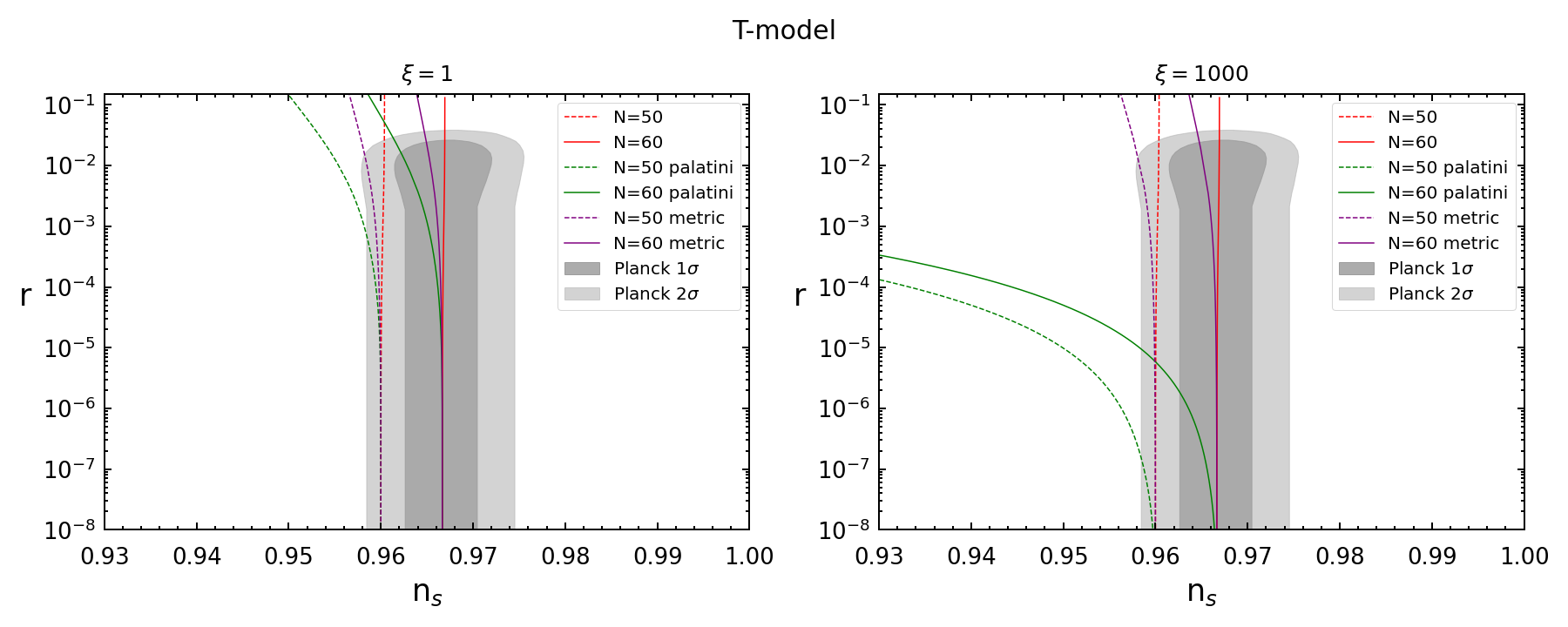}
\end{center}
\caption{$r$ vs. $n_s$ for the alpha attractor potential in eq. \eqref{eq:attractor} with  $-6 < \log_{10} \alpha < 4$ computed in the Einstein frame (red), in the metric Jordan frame (purple) and in the Palatini Jordan frame (green). Dashed (continuous) line represents $N_e=50\ (60)$. The gray areas represent the $1,2\sigma$ allowed regions from the latest combination of Planck, BICEP/Keck and BAO data \cite{BICEP:2021xfz}. 
}
\label{fig:alpha}
\end{figure}
We consider two different Jordan frames, one corresponding to a metric formulation of gravity (see eq. \eqref{eq:dphim}) and one to the Palatini one (see eq. \eqref{eq:dphiP}). The results for $r$ vs. $n_s$ are presented in Figs. \ref{fig:powerlaw}-\ref{fig:alpha} for 50 (dashed line) and 60 (continuous line) $e$-folds, computed in the Einstein frame (red), in the metric Jordan frame (purple) and in the Palatini Jordan frame (green). The gray areas represent the $1,2\sigma$ allowed regions from the latest combination of Planck, BICEP/Keck and BAO data \cite{BICEP:2021xfz}. On the left panels we show the results for $\xi=1$ and on the right ones for $\xi=1000$.
In Fig. \ref{fig:powerlaw} we have the results for the power-law potential in eq.  \eqref{eq:powerlaw} with $-2 < \log_{10}(n) < 2$.
In Fig. \ref{fig:natural} we have the results for the natural inflation potential in eq.  \eqref{eq:natural} with $0 < \log_{10}(\mu/M_p) < 2.5$.
In Fig. \ref{fig:Hilltop} we have the results for the quartic hilltop potential in eq.  \eqref{eq:hilltop} with $-2<\log_{10}(\mu/M_p) < 2$.
%
%
In Fig. \ref{fig:alpha} we have the results for the $\alpha$-attractor T-model in eq. \eqref{eq:attractor} with $-6 < \log_{10}(\alpha) <4$.
In all the cases we can see that the difference in the results increases with $\xi$ increasing. However the increase in the metric Jordan frame is more or less contained: between $\xi=1$ and $\xi=1000$ the difference is not that remarkable but still visible in the plots. 
On the other hand, the increase of the discrepancy from $\xi=1$ to $\xi=1000$ in the Palatini Jordan frame is dramatic. While for $\xi=1$ the predictions in the Palatini Jordan frame are usually only slightly altered, for $\xi=1000$ they are completely changed. 
In the power law case, Fig. \ref{fig:powerlaw}, from being ruled out in all the regions of the parameters space because of a too large $r$ or $n_s$, we pass to a strong reduction of $r$ and $n_s$   values (even reaching the core of 1$\sigma$ region) and being allowed for very small values of $n$. 
In the natural inflation case, Fig. \ref{fig:natural}, from being ruled out in all the region of the parameters space because of a too large $r$ or too small $n_s$ for $\xi=1$, we again completely invert the situation for $\xi=1000$ with $r$ and $n_s$ values even in the core of 1$\sigma$ region for $5.18 < \mu/M_p < 5.82$
In the quartic hilltop inflation case, Fig. \ref{fig:Hilltop},  for $\xi$=1 the allowed region is $6.1 < \mu/M_p^2 < 35.5$ with $r\sim 10^{[-3,-2]}$. Instead for $\xi=1000$ it is $0.68 < \mu/M_p^2 < 2.85$ with $r\sim 10^{[-6,-5]}$.
%
%
In the $\alpha$-attractor case, Fig. \ref{fig:alpha},  for $\xi$=1 the discrepancy is visible only for  $r\sim 10^{[-3,-2]}$ and most of the parameters space is in the allowed region. Instead for $\xi=1000$, both $r$ and $n_s$ values decrease sensibly and a lot of the parameters space is out of the allowed region.

\section{A possible solution} \label{sec:solution}
The number of $e$-folds describes how much the Universe expanded because of inflation and it is usually defined directly from the scale factor of the metric: 
\be
 N_e = \ln\left[ \frac{a(t_f)}{a(t_N)} \right] \label{eq:Ne:old} \, .
\ee
An alternative but more operative definition would involve instead the physical distance between two fixed points in the comoving frame measured at the times $t_N$ and $t_f$
\be
N_e = \ln\left[ \frac{\Delta {\bf r}(t_f)}{\Delta {\bf r}(t_N)} \right] = \ln\left[ \frac{a(t_f) \Delta {\bf x}}{a(t_N) \Delta {\bf x}} \right] \equiv \ln\left[ \frac{a(t_f)}{a(t_N)} \right] \label{eq:Ne:other}
\ee
where $\Delta {\bf r}$ is the physical distance and $\Delta {\bf x}$ is the comoving distance.
The definitions \eqref{eq:Ne:old} and \eqref{eq:Ne:other} are exactly equivalent and from both we can immediately see the issue with frame invariance. The metric (and therefore the scale factor) and absolute physical distances are not frame invariant and therefore neither is the number of $e$-folds. On the other hand, operatively, we do not measure absolute distances, but only relative ones i.e. we need a unit of measurement. Therefore, as customary (e.g. \cite{Postma:2014vaa}), we introduce the concept of \emph{normalized distance} i.e. the ratio of the distance over the fundamental reference length, the Planck length\footnote{Historically the Planck mass $M_P$ and length $\ell_P$ have been derived as combinations of three fundamental constants: the Planck constant $h$, the speed of light $c$ and Newton's constant of gravitation $G_N$. However, once we accept General Relativity as the proper description of gravity and Newtonian gravity as an effective description, we implicitly assume $M_P$ (or $\ell_P$) to be the fundamental constant and $G_N$ to be the derived one. $h$ and $c$ are frame invariant (it can be explicitly seen in natural units, where they are dimensionless), while $M_P$ ($\ell_P$) and $G_N$ are frame dependent. This is fine since gravity looks different in each frame.}  of the corresponding frame: $\ell_P^i$, where $i$ labels the frame. Hence, we can define the normalized physical distance as 
\be 
\Delta {\bf \bar r} = \frac{\Delta {\bf r_i}}{\ell_P^i} = \frac{a(t)_i \Delta {\bf x}}{\ell_P^i}
\ee
where $i$ labels the frame. It is easy to verify that such a definition is actually frame invariant. In the Einstein frame we have
\be 
\Delta {\bf \bar r} = \frac{\Delta {\bf r_E}}{\ell_P^E} =  a(t)_E \, \Delta {\bf x} \, M_P \label{eq:rbar:E}
\ee
while in the Jordan frame we have
\be 
\Delta {\bf \bar r} = \frac{\Delta {\bf r_J}}{\ell_P^J} =  a(t)_J \, \Delta {\bf x} \, \sqrt{f(\phi)} M_P = a(t)_E \, \Delta {\bf x} \, M_P \label{eq:rbar:J}
\ee
where we used eq. \eqref{eq:tE:aE}. The Jordan frame normalized distance in eq. \eqref{eq:rbar:J} is identical to the Einstein frame one in eq. \eqref{eq:rbar:E}. Therefore we can define an invariant number of $e$-folds
\be
 \bar N_e \equiv \ln\left[ \frac{\Delta {\bf \bar r}(t_f)}{\Delta {\bf \bar  r}(t_N)} \right]
 \label{eq:Ne:inv}
\ee
We can easily check that such a definition recovers the standard Einstein frame definition in both the Jordan and Einstein frame
\bea
 \bar N_e &=& \ln\left[ \frac{\Delta {\bf \bar r}(t_f)}{\Delta {\bf \bar  r}(t_N)} \right] 
 \label{eq:Ne:E}\\ 
 &=&
 \ln\left[ \frac{\Delta {\bf  r_J}(t_f)/\ell_P^J}{\Delta {\bf   r_J}(t_N)/\ell_P^J} \right] =
 \ln\left[ \frac{a(t_f)_E \, \Delta {\bf x} \, M_P}{a(t_N)_E \, \Delta {\bf x} \, M_P} \right]  = \ln\left[ \frac{a(t_f)_E}{a(t_N)_E} \right] \nn\\
 &=& 
 \ln\left[ \frac{\Delta {\bf  r_E}(t_f)/\ell_P^E}{\Delta {\bf   r_E}(t_N)/\ell_P^E} \right]=
  \ln\left[ \frac{a(t_f)_E \, \Delta {\bf x} \, M_P}{a(t_N)_E \, \Delta {\bf x} \, M_P} \right] =
 \ln\left[ \frac{a(t_f)_E}{a(t_N)_E} \right] \nn
\eea
solving the discrepancy between the old standard definitions of the number of $e$-folds in different frames\footnote{The careful reader might notice that the invariant number of $e-$folds in eq. \eqref{eq:Ne:inv} turns out to be the same as the regular number of $e-$folds computed in the Einstein frame (cf. eqs. \eqref{eq:Ne:old} and \eqref{eq:Ne:E}). First of all we stress that this is not anymore a definition but the final mathematical outcome coming from our frame invariant definition (i.e. valid in both frames) in eq. \eqref{eq:Ne:inv}. Second, such a result does not give the Einstein frame a special status in the \emph{physical} sense, because, when properly defined, inflationary observables are frame invariant (e.g \cite{Chiba:2008ia,Chiba:2008rp,Jarv:2016sow}). On the other hand, we might say that the Einstein frame is \emph{computationally} special, being the only frame where $f=1$ (cf. eqs. \eqref{eq:rbar:E} and \eqref{eq:rbar:J}) and therefore computations the easiest (e.g \cite{Chiba:2008ia,Chiba:2008rp,Jarv:2016sow}).}. To conclude we stress that according to our knowledge, even though the notion of normalized distance is not novel, its application to the definition of the number of $e$-folds is.

\section{Conclusions} \label{sec:conclusions}
In this article we investigated how inflationary predictions are affected by the difference in the number of $e$-folds between the Jordan and Einstein frames. We studied as test examples four well-known models: monomial inflation, natural inflation, quartic hilltop inflation and $\alpha$-attractors. We considered a Jordan frame defined by the popular Higgs-inflation-like non-minimal coupling to gravity and  two different formulations of gravity: metric and Palatini. We found that for all tested models, the difference between the predictions for an Einstein frame number of $e$-folds and a metric Jordan frame one remains quite contained regardless of the size of the non-minimal coupling to gravity, but still appreciable in the $r$ vs. $n_s$ plot. On the other hand the difference between the results for an Einstein frame number of $e$-folds and a Palatini Jordan frame one becomes gigantic in case of a relatively big non-minimal coupling to gravity. This might have a tremendous impact in ruling in or out inflationary models, specially in light of the precision of the forthcoming experiments (e.g. Simons Observatory \cite{SimonsObservatory:2018koc}, PICO \cite{NASAPICO:2019thw}, CMB-S4 \cite{Abazajian:2019eic} and LITEBIRD \cite{LiteBIRD:2020khw}). 

Finally, we discuss a way to overcome the discrepancy by using the concept of \emph{normalized distance}, i.e. the ratio of the distance to the Planck length in the corresponding frame. It is well known that the normalized physical distance is frame invariant. In particular it can be used to define an invariant number of $e$-folds as the ratio of the normalized physical distance (between two fixed points in the comoving frame) evaluated at end of inflation to the same distance at the time when a given scale leaves the horizon. According to our knowledge, even though the notion of normalized distance is not novel, such an application is. If accepted, this new definition of the number of $e$-folds, would put an end to the ambiguity of inflationary computations between the Jordan and Einstein frame.

\section*{Acknowledgments}
The authors thanks Kristjan Kannike, Luca Marzola and Marco Piva for useful discussions.
This work was supported by the Estonian Research Council grants MOBTT5, MOBTT86, PRG803, PRG1055
and by the EU through the European Regional Development Fund
CoE program TK133 ``The Dark Side of the Universe''. 

\appendix 
\section{More details on the formulations of gravity} \label{appendix}
In this appendix we give more details about the formulations of gravity (metric or Palatini) and the corresponding transformations that lead to the Einstein frame. In order to have a more readable Appendix, we will repeat some of the equations already used in the previous sections.
Our starting point is the action in eq. \eqref{eq:JframeL}
\begin{equation}
S = \!\! \int \!\! d^4x \sqrt{-g^J}\left(-\frac{M_P^2}{2}f(\phi)R^J(\Gamma) + \frac{(\partial \phi)^2}{2}  - V(\phi) \right) ,
\label{eq:JframeL:app}
\end{equation}
where $M_P$ is the reduced Planck mass, $R^J$ is the Jordan frame Ricci scalar constructed from a connection $\Gamma$, $f(\phi)$ is the non-minimal coupling to gravity and $V(\phi)$ is the potential of the inflaton scalar. 
In order to avoid repulsive gravity we require $f(\phi) > 0$. 
This feature is independent on the eventual gravity formulation (metric or Palatini).
In the usual metric formulation the connection is uniquely determined as a function of the metric tensor, i.e.\ it is the Levi-Civita connection $\bar{\Gamma}=\bar{\Gamma}(g^{\mu\nu})$
\begin{eqnarray}
\label{eq:LC:app}
\overline{\Gamma}^{\lambda}_{\alpha \beta} = \frac{1}{2} (g^J)^{\lambda \rho} \left( \partial_{\alpha} g^J_{\beta \rho}
+ \partial_{\beta} g^J_{\rho \alpha} - \partial_{\rho} g^J_{\alpha \beta}\right) \, .
\end{eqnarray}
On the other hand, in the Palatini formalism both $g^J_{\mu\nu}$ and $\Gamma$ are considered independent variables, and the only assumption is that the connection is torsion-free, $\Gamma^\lambda_{\alpha\beta}=\Gamma^\lambda_{\beta\alpha}$. Solving the equations of motion leads to \cite{Bauer:2008zj}
\begin{eqnarray}
\Gamma^{\lambda}_{\alpha \beta} = \overline{\Gamma}^{\lambda}_{\alpha \beta}
+ \delta^{\lambda}_{\alpha} \partial_{\beta} \omega(\phi) +
\delta^{\lambda}_{\beta} \partial_{\alpha} \omega(\phi) - g_{\alpha \beta} \partial^{\lambda}  \omega(\phi) \, ,
\label{eq:conn:J:app}
\end{eqnarray}
where
\begin{eqnarray}
\label{omega:app}
\omega\left(\phi\right)=\ln\sqrt{f(\phi)} \, .
\end{eqnarray}
Therefore the connection in the Palatini formulation (eq. \eqref{eq:conn:J:app}) is the same as in the metric formulation (the Levi-Civita connection in eq. \eqref{eq:LC:app}) plus a correction that depends on the on minimal coupling $f(\phi)$ via the function $\omega(\phi)$ in eq. \eqref{omega:app}.
Because the connections (\ref{eq:LC:app}) and (\ref{eq:conn:J:app}) different, the metric and Palatini formulation provide indeed two different theories of gravity when $f(\phi) \neq 1$.
Another way of seeing the differences is to investigate the problem in the Einstein frame by means of the Weyl transformation
\begin{eqnarray}
\label{eq:gE:app}
g_{\mu \nu}^E = f(\phi) \ g^J_{\mu \nu} \, .
\end{eqnarray}
In the Einstein frame gravity looks the same in both the formalisms (see also eq. (\ref{eq:conn:J:app})), however the matter sector (in our case $\phi$) behaves differently. Performing the computations \cite{Bauer:2008zj} in both formulations, the Einstein frame action can be formally expressed as
\begin{equation}
S = \!\! \int \!\! d^4x \sqrt{-g^E}\left(-\frac{M_P^2}{2} R^E + \frac{(\partial \chi)^2}{2}  - U(\chi) \right) ,
\label{eq:L:E:app}
\end{equation}
where $\chi$ is the canonically normalized scalar field in the Einstein frame, and its scalar potential is
\be
U(\chi) = \frac{V(\phi(\chi))}{f^{2}(\phi(\chi))} \, .
\label{eq:U:app}
\ee
In case of the metric formulation, $\chi$ is derived by integrating the following relation
\begin{equation}
\frac{\partial \chi}{\partial \phi} = \sqrt{\frac{3}{2}\left(\frac{M_P}{f}\frac{\partial f}{\partial \phi}\right)^2+\frac{1}{f}} \, ,  
  \label{eq:dphiE:app}
\end{equation}
where, as well known (e.g. \cite{Bezrukov:2013fka}), the first term comes from the transformation of the Jordan frame Ricci scalar and the second from the rescaling of the Jordan frame scalar field kinetic term. 

On the other hand, in the Palatini formulation, while the Einstein frame scalar potential remains eq. \eqref{eq:U:app}, the field redefinition changes ( e.g. \cite{Bauer:2008zj}). In order to understand better the change, let us remind how a Weyl transformation acts on the curvature tensors in case of Palatini gravity. In such a case the connection $\Gamma^\lambda_{\rho\sigma}$ and the metric  $g^J_{\mu\nu}$ are independent variables, therefore the Riemann tensor $(R^J)^{\lambda}_{\ \, \mu\nu\sigma}(\Gamma, \partial\Gamma)$, being constructed from $\Gamma$ and its first derivatives, is invariant under any transformation of the sole metric. The same holds for the Ricci tensor which is built by contraction of the Riemann tensor with a Kronecker delta:  $R^J_{\mu\nu}(\Gamma, \partial\Gamma)={\delta^\nu}_\lambda (R^J)^{\lambda}_{\ \, \mu\nu\sigma}(\Gamma, \partial\Gamma)$. On the other hand, the curvature (Ricci) scalar $R^J=(g^J)^{\mu\nu} R^J_{\mu\nu}(\Gamma, \partial\Gamma)$ is explicitly dependent on the metric and therefore under the rescaling \eqref{eq:gE:app}
$R$ scales inversely,
\be
   R^E = \frac{R^J}{f(\phi)} \label{eq:R:scaling:app} \, .
\ee
Therefore it is easy to check that the action term linear in $R^J$ in eq. \eqref{eq:JframeL:app} transforms into the Einstein-Hilbert action with no additional term contributing to the kinetic term of $\phi$. Therefore, the Einstein frame  scalar field redefinition is now induced only by the rescaling of the inflaton kinetic term i.e.
\begin{equation}
\frac{\partial \chi}{\partial \phi} = \sqrt{\frac{1}{f}} \, . 
\end{equation}
Therefore we can see that the difference between the two formulations in the Einstein frame relies on the different definition of $\chi$ induced by the different non-minimal kinetic term involving $\phi$.

\bibliography{references}

\end{document}